\documentclass[preprint,eqsecnum,aps]{revtex4}

\usepackage{graphicx}
\usepackage{latexsym}
\usepackage{amsmath}
\usepackage{amssymb}

\newcommand{\beq}{\begin{equation}}
\newcommand{\eeq}{\end{equation}}
\newcommand{\bea}{\begin{eqnarray}}
\newcommand{\eea}{\end{eqnarray}}
\newcommand{\bfig}{\begin{figure}}
\newcommand{\efig}{\end{figure}}
\newcommand{\nno}{\nonumber}

\begin{document}
\title{\Large \bf Modulus stabilization of generalized Randall Sundrum model with bulk scalar field}
\author{Ratna Koley${}^{}$ \footnote{E-mail:tprk@iacs.res.in}, 
Joydip Mitra${}^{}$ \footnote{E-mail: tpjm@iacs.res.in} and  
Soumitra SenGupta${}^{}$ \footnote{E-mail: tpssg@iacs.res.in}}
\affiliation{ Department of Theoretical Physics and Centre for
Theoretical Sciences,\\
Indian Association for the Cultivation of Science,\\
Kolkata - 700 032, India}

\begin{abstract}
We study the stabilization of inter-brane spacing modulus of generalized warped brane models 
with a nonzero brane cosmological constant. Employing Goldberger-Wise stabilization prescription of brane world models
with a bulk scalar field,
we show that the stabilized value of the modulus generally  depends on the value of the brane cosmological constant.
Our result further reveals that the stabilized modulus value corresponding to a vanishingly small cosmological constant
can only resolve the gauge hierarchy problem simultaneously. This in turn 
vindicates the original Randall-Sundrum model where the 3-brane cosmological constant was 
chosen to be zero.

\end{abstract}

\maketitle

\section{Introduction}

A novel solution to 
the gauge hierarchy problem in the standard model of elementary particles 
was proposed by Randall and Sundrum  \cite{RS} through an exponential 
warping of the space-time metric in a five dimensional anti-de Sitter space-time 
compactified on a  $S^1/Z_2$ orbifold. The five dimensional metric has a solution:
\beq
ds^2 = e^{- k r_c  \vert \phi \vert} \eta_{\mu \nu} dx^{\mu} dx^{\nu} + r^2_c d \phi^2
\eeq

where $k$ is of the order of Planck scale ($M_{Pl}$) and $r_c$ specifies the radius of the extra 
dimension. Two 3-branes are located at the orbifold fixed points $\phi = \pi$ (visible) 
and $\phi = 0$ (hidden). Due to warping 
of 4D space-time metric all the fundamental scalar masses get rescaled on the visible brane and  
a fundamental mass ($m_0$) of Planck scale order comes down to a physical mass scale 
($m$) of order TeV on the visible brane through the relation :  $m = m_0 e^{-k r_c \pi}$. 
In the RS model the  radius of extra dimension is suitably {\em fine tuned} 
as $k r_c \approx 12$ so that the gauge hierarchy problem can be resolved without introducing any 
intermediate mass scale in the theory.
This can be achieved by choosing $k$ ( which is related to the bulk cosmological constant) 
and the inverse of $r_c$ of the order of Planck mass. 
However without any dynamical basis for this fine tuning, 
it became imperative to introduce dynamics which determines a stabilized location of the 
TeV brane relative to the Planck brane. 

It was proposed by Goldberger and Wise (GW) \cite{gw} that the dynamics of a 
five dimensional bulk scalar field in such models could stabilize the size of the extra dimension
to it's desired value. In this mechanism the effective potential 
for the modulus has been generated by the total action integral of a bulk scalar 
field with quartic interactions localized in two 3-branes. The 
bulk scalar then acts like a spring between the branes, whose gradient energy 
becomes repulsive if the branes get too close, and whose potential energy 
(from $m^2 \phi^2$) causes attraction if the branes separate too much. 
In this way the minimum of the modulus ( brane separation) potential yields a compactification scale in 
terms of the ratio of VEV's of the scalar fields at the two branes which solves the gauge hierarchy problem 
{\em without fine tuning any of the parameter of the theory}. 
Initially GW neglected 
the back reaction of the bulk scalar on the geometry as well as on the 
brane tensions. Later the back reaction has been included in \cite{dewolf} and 
some critical studies have been done in \cite{gwssg}. Several variations 
of the GW mechanism in higher dimensions \cite{gw6d}, in cosmological 
braneworlds \cite{gwcos} and with non minimal scalar fields \cite{gwfol} 
have been studied rigorously.  This stabilization mechanism has also 
been generalized for supersymmetric warped space-times \cite{gwfol}.  

In the present work we will apply the  GW mechanism of radius stabilization 
in the braneworld
model with nonzero brane cosmological constant. As can be seen that the  original Randall-Sundrum model 
started with  flat 3-branes, it has been shown in \cite{ssd} that one 
can indeed generalize the model with non-zero cosmological 
constant on the brane and still can have Planck to TeV scale warping from the 
resulting warp factor. Different values of the inter-brane separation modulus  
for different values for the  brane cosmological constant ($\Omega$) are obtained through their functional relation
such that the gauge hierarchy 
problem is resolved for this entire region of the parameter space. The analysis has been done for both 
de-Siiter and 
anti-de Sitter TeV branes by choosing appropriate signature for $\Omega$  and the corresponding
brane tensions have been determined. Motivations for this analysis as well as that presented in this work 
have their roots in the observational support in favour of de-Sitter-like character of our universe. The present accelerating phase
of our universe can be explained by the presence of a small but positive cosmological constant whose value is estimated
to be $\sim 10^{-124}$ in Planck unit. Despite such a tiny value it is responsible in inducting such a large acceleration
to the expansion of the universe. It is therefore worthwhile to to explore the effects of a non-vanishing cosmological constant on the
brane particularly in the context of gauge hierarchy and modulus stabilization.\\ 
In this work we study the 
stabilization of such a generalized RS model in the light of Goldberger-Wise mechanism
by introducing a scalar field in the bulk. We show that such a stabilization is indeed possible and 
the stable value of the modulus depends on the value of the brane cosmological constant.
However for a given value of the cosmological constant the corresponding stable value of the modulus will resolve the
gauge hierarchy problem also only when
the value of the cosmological constant is vanishingly small. In other words the modulus-cosmological constant
relation to resolve the gauge hierarchy problem and the modulus -cosmological relation to
achieve stability will have common solution only in the region of vanishingly small cosmological constant.
We shall discuss these issues more quantitatively
in the context of the presently observed value of the cosmological constant at the end of our analysis.

In this paper we first briefly summarize the braneworld model with non zero 
brane cosmological constants as found in \cite{ssd}. The detailed analysis of the stability mechanism has been 
presented in the next section. Finally we conclude with the summary of our results 
and discuss some open issues. 

\section{Model}

In the Randall and Sundrum \cite{RS} warped braneworld model the cosmological 
constant induced on the visible brane is zero which is not consistent with the present day  
observed small positive value of the cosmological constant. Moreover the consistent boundary condition
requires that the tension of the visible TeV brane must be 
negative which in turn indicates an intrinsic instability. 
In an attempt to generalize this model it has been demonstrated in \cite{ssd} that the condition of zero
cosmological constant on the brane  
may be relaxed for a more general warp factor which 
includes branes with non-zero cosmological constant. Another important 
aspect of the model \cite{ssd} is that in some circumstances the visible 
brane tension may become positive also. Let us now discuss the model briefly.  
In this model the warped 
geometry has been obtained by considering a constant curvature brane space-time.  
General form of the warped metric for a five dimensional space-time is given by, 
\bea
\label{met}
ds^2 = e^{- 2 A(y)} g_{\mu\nu} dx^{\mu} dx^{\nu} + dy^2 \label{metric}
\eea
where $g_{\mu\nu}$ stands for four dimensional curved brane. There are two branes
located at orbifold fixed points $y = 0$ and $ y = r \pi$ where $r$ corresponds 
to the modulus associated with the extra dimension. The bulk consists of a 
negative cosmological constant, $\Lambda$. A scalar mass 
on the visible brane gets warped through the warp factor $e^{-A(k r \pi)}=\frac{m}{m_0}=10^{-n}$ 
where, $k = \sqrt{- \frac{\Lambda}{12 M^3}}$ $\sim$ Planck 
Mass and `$n$', the warp factor index is set to $16$ to achieve the desired warping from Planck to
TeV scale. 
The magnitude of the induced cosmological constant on the brane in this case is 
non-vanishing in general and is given by = $10^{-N}$ (in Planck units). For a  
negative brane cosmological constant, N has minimum value given by $N_{min}=2 n$ 
leading to an upper bound on the magnitude of the cosmological constant ( depending on the choice
of the warp factor index $n$. 
There is no such upper bound for the induced positive cosmological constant on the brane.
For brane cosmological constant, $\Omega > 0$ and $\Omega < 0$, the 
brane metric $g_{\mu\nu}$ may correspond to different space-time geometry  such as dS-Schwarzschild and AdS-Schwarzschild 
spacetimes respectively.

\subsection{AdS Brane :}
For the negative value of the cosmological constant $\Omega$ on the visible brane if one redefines 
$\omega^2 \equiv -\Omega/3k^2 \geq 0$ then the solution of the warp factor can be written as :
\bea 
\label{wfads}
e^{-A} = \omega \cosh\left(\ln \frac {\omega} c_1 + ky \right)
\eea
where $c_1 = 1 + \sqrt{1 - \omega^2}$ for which the warp factor is normalized to unity at $y = 0$. 
One can show that real solution for the warp factor exists if and only 
if $\omega^2 \leq 10^{-2n}$. This leads to an upper bound for the magnitude 
of the cosmological constant as $N_{min}=2 n$. So, for $n=16$ , $N$ is 
found to be $10^{-32}$. For $N=N_{min}$, one obtains a degenerate solution~
 $k r \pi = n \ln {10}+ \ln{2}$. For $N-2n>>1$, the solutions obtained in this case, are
\bea
k r_1 \pi & = & n \ln{10}+\frac{1}{4}10^{-(N-2n)}  \nno \\
k r_2 \pi & = & (N-n)\ln{10}+\ln{4} 
\eea
As an example, to have a solution of the gauge hierarchy problem with the small non zero 
value of the brane cosmological constant we consider $n=16$ and $N=124$ for 
which the values in the above equation become 
\bea 
k \pi r_1 & \simeq & 36.84 + 10^{-93} \nno \\
k \pi r_2 & = & 250.07 
\eea

Note that for $x=n \ln{10}$ and $N=\infty$, the RS value can be recovered. 
At $k r \pi = n \ln{10}+\ln{2}$ and $N=2 n$, $\omega^2$ reaches to its maximum value. 
Beyond this the magnitude of $\omega^2$ starts to decrease again. 
One can also obtain the tension of the visible brane for the above two solutions. 
When $N=N_{min}=2 n$,  the  visible brane tension is zero.
For $r = r_1$ it is negative while for $r = r_2$ it is positive.

\subsection{dS Brane :}

For the induced brane cosmological constant $\Omega > 0$ the warp factor turns out to be :
\bea
\label{wfds}
 e^{-A} = \omega \sinh\left(\ln \frac {c_2}{\omega} - ky \right)~,
\eea
where  $\omega^2 \equiv \Omega/3k^2$ and $ c_2 = {1 + \sqrt{1
+ {\omega}^2}}~.$  In this case there is no bound on the value of $\omega^2$, 
and the (positive) cosmological constant can be of arbitrary magnitude. 
Also, there is a single solution of $k r \pi$ whose precise value will depend 
on $\omega^2$ and $n$. The brane tension is negative for the entire range of values of 
the positive cosmological constant. In Fig.(1) it has been 
shown how the value of $\omega^2$ is related to the modulus which solves the hierarchy problem.

Moreover it has been shown in \cite{ssd} that if one wants to resolve the fine 
tuning problem in connection with Higgs mass without introducing any intermediate
scale through the modulus $r$ ( i.e keeping it close to Planck length) or the 
parameter $k$ ( keeping it close to Planck mass) then the cosmological 
constant must be tuned to a very tiny value. Therefore the issues of gauge hierarchy problem and the 
cosmological constant problem are intimately related to each other. An interesting outcome of this generalised RS model has been found
in the context of localization of fermions on the Tev brane \cite{rkjmssg}. It has been shown that although
the generalised RS model allows different values of $kr$ for different values of the cosmological 
constant ( both in dS and AdS region in the figure above ) for the resolution of the hierarchy problem but 
the simultaneous requirement of fermion localization can be achieved only for vanishingly small cosmological constant i.e. 
when the warp factor is close to that obtained in the original RS model.

With this generalized RS braneworld model we now explore the mechanism of stabilizing the 
modulus $r$ to a desired value ( which of course depend on the value the cosmological
constant $\omega$ ) so that the desired Planck to TeV scale warping can be achieved.
 \begin{figure} 
\includegraphics[width=3.330in,height=2.20in]{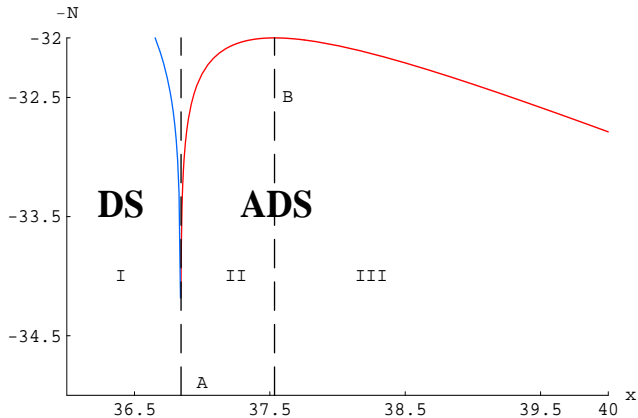}
\caption{Graph of $N$ versus $x = k r \pi =36-40$, for $n=16$ and for both
positive and negative brane cosmological constant. The curve in
region-I corresponds to positive cosmological constant on the brane,
whereas the curve in regions-II \& III represents negative
cosmological constant on the brane. } \label{ADSDS}
\end{figure}

\section{Stability Mechanism}
To stabilize the modulus $r$ we adopt the method proposed by Goldberger and Wise \cite{gw} in the context of the original RS model.
Let us consider a scalar field, $\Phi$ in the bulk with quartic interactions 
localized in the visible ($y = r \pi$) and Planck ($y = 0$) brane for which 
the bulk action turns out to be :
\beq 
\label{scalarac}
S = \frac{1}{2} \int d^4 x \int_0^{r \pi} dy (G^{AB}\partial_A \Phi \partial_B 
\Phi -m^2 \Phi^2) + S_h~ \delta(y) + S_v ~\delta(y - r \pi)
\eeq 

where $G_{AB}$ with $A,B$ = $\mu$, $y$ is given by the metric (\ref{met}) and 
the interaction terms on the hidden and visible branes are given by 
\bea
S_h = -\int d^4 x \sqrt{-g_h} \lambda_h (\Phi^2-v_h^2)^2 \\
S_v = -\int d^4 x \sqrt{-g_v} \lambda_v (\Phi^2-v_v^2)^2
\eea
Here $g_h$ and $g_v$ are the determinants of the induced metric on the hidden 
and visible brane respectively. The vacuum  expectation values are given by 
$v_h$ and $v_v$ which have mass dimension 3/2 and $\lambda_h$ and  $ \lambda_v$ stands for the 
brane tensions.  For simplicity we assume that the scalar field is a function of the 
extra dimension $y$ only. The equation of motion satisfied by the field $\Phi(y)$ 
is given by,
\beq
\partial_y(e^{-4 A(y)} \partial_y \Phi)- e^{-4 A(y)} \left( m^2 + 4 \lambda_v 
\Phi (\Phi^2-v_v^2) \delta(y- r \pi) + 4 \lambda_h \Phi (\Phi^2-v_h^2)\delta(y) \right) = 0
\eeq

Now we study the nature of the solution of the scalar field for the warp factor 
given in (\ref{wfads}). It has been discussed in the previous section that to resolve 
the gauge hierarchy problem without introducing any intermediate scale, the brane cosmological 
constant must be tuned to a very small value. Moreover the present observed value of the cosmological constant
is also extremely tiny. We therefore consider small $\omega$ limit 
to solve the scalar field equation. 

In this case the warp factor in (\ref{wfads}) takes the following perturbative form
\beq
e^{-4 A(y)}=e^{-4 k y}+\omega^2 e^{-2 k y}
\eeq

With the above warp factor the scalar field equation away from the brane locations (i.e. inside the bulk) 
reduces to the following form: 
\beq
\partial_y^2 \Phi-4 k \partial_y \Phi +2 k \omega^2 e^{2 k y}+m^2 \Phi=0
\eeq

Solving the above equation we obtain the solution for the scalar field as, 
\bea
\label{scalar}
\Phi(y) &=& e^{2 k y}\left[A e^{\nu k y}+B e^{-\nu k y}\right] \nno \\ 
& -& \frac{\omega^2}{2} \left[\tilde A\left(\frac{\omega}{\sqrt{2}} \right)^{\nu}\left(\frac{\nu+2}{\nu+1}\right) e^{(\nu + 4)k y} + \tilde B \left(\frac{\omega}{\sqrt{2}} \right)^{\nu} \left(\frac{-\nu+2}{-\nu+1}\right)e^{(-\nu+4)k y}\right]
\eea
where $\nu=\sqrt{4+m^2/k^2}$.\\ 
In the above solution we have 
defined $A $ = $\tilde A \left(\frac{\omega}{\sqrt{2}} \right)^{ \nu}$ and $B $ = $\tilde B \left( \frac{\omega}{\sqrt{2}} \right)^{ \nu}$ where $\tilde A$ and $\tilde B$ are arbitrary constants. 
The constants can be evaluated by using the appropriate boundary conditions at the locations of the branes. 

An effective potential ($V_{eff}$) for the modulus $r$ can now be obtained by putting the above 
solution (\ref{scalar}) back into the scalar field action and 
integrating the entire scalar field action (\ref{scalarac}) over the extra dimension \cite{gw}.
This yields an effective modulus potential as, 
\bea
\label{effpot}
V_{eff} &=& k A^2(\nu+2)(e^{2 \nu k r \pi}-1)+k B^2(2-\nu)(e^{-2 \nu k r \pi}-1) \nno \\
&+& k \omega^2 [2 A^2 \left(\frac{\nu+2}{\nu+1}\right)(1-e^{(2 \nu +2)k r \pi}) \nno \\ &+& 2 B^2 \left(\frac{2-\nu}{1-\nu}\right)(1-e^{(2-2 \nu)k r \pi})+AB \frac{\nu}{1-\nu^2}(e^{2 k r  \pi}-1)] \nno \\ 
&+& \lambda_v e^{-4 A_{\pi}}(\Phi(\pi)^2-v_v^2)+\lambda_h (\Phi(0)^2-v_h^2)
\eea
We now calculate the unknown coefficients $A$ and $B$. Rather than solving the 
equations in general (i.e. putting the solution back in the action and matching the delta functions ) 
we consider the simplified case where the parameters $\lambda_v$ and $\lambda_h$ are large \cite{gw}. It is seen from 
the expression (\ref{effpot}) that in this limit it is energetically favorable to have $\Phi(0)$ = $v_h$ and $\Phi(r \pi)$ = $v_v$. 
For large $k r$ limit the expressions for $A$ and $B$ turn out to be,
\bea
A&=&v_v e^{(2+\nu) k r \pi}-v_h e^{-2 \nu k r \pi} \nno \\ &+& \frac{\omega^2}{2} v_h\left[\left(\frac{2 \nu}{1-\nu^2}-\frac{2+\nu}{1+\nu}\right)e^{-2 \nu k r \pi}+\frac{2(2-\nu^2)}{1-\nu^2}e^{(2-2 \nu)k r \pi}\right] \nno \\
&+&\frac{\omega^2}{2} v_v\left[\left(\frac{2+\nu}{1+\nu}\right)e^{-\nu k r \pi}\right]
\eea
 and 
\bea
B&=& v_h\left(1+e^{-2 \nu k r \pi}\right)-v_v e^{-(2+\nu)k r \pi} \nno \\ &+&\frac{\omega^2}{2} v_h\left[\left(\frac{2 \nu}{1-\nu^2}\right)+\left(\frac{2+\nu}{1+\nu}\right)\left(1+e^{-2 \nu k r \pi}\right)-\frac{2(2-\nu^2)}{1-\nu^2}e^{(2-2 \nu)k r \pi}\right]\nno \\ &-&\frac{\omega^2}{2} v_v\left[\left(\frac{2 \nu}{1-\nu^2}\right)e^{-(\nu+2)k r \pi}+\left(\frac{2+\nu}{1+\nu}\right)e^{-\nu k r \pi}\right]
\eea
where subleading powers of $e^{-k r \pi}$ have been neglected. Now suppose that $m/k\ll1$
so that $\nu=2+\epsilon$ , with $\epsilon \simeq m^2/4 k^2$ is a small quantity. In the large $k r$ limit,the 
expression of the potential becomes,
\bea
V_{eff}&=&(4+\epsilon)e^{-4 k r \pi}(v_v-v_h e^{-\epsilon k r \pi})^2+(4+\epsilon)e^{-4 k r \pi}2 \omega^2 (v_v-v_h e^{-\epsilon k r \pi})(v_v F(k)+v_h e^{-\epsilon k r \pi} G(k)) \nonumber \\
&-&\omega^2 \frac{2(4+\epsilon)}{3+\epsilon}(v_v-v_h e^{-\epsilon k r \pi})^2 e^{-2 k r \pi}+\epsilon \left[v_h^2(1+e^{-2(2+\epsilon) k r \pi})-2 v_v v_h e^{-(4+\epsilon)k r \pi}\right] \nonumber \\
&+& 2 \omega^2 \epsilon  \left[v_h^2 (1+e^{-2(2+\epsilon)k r \pi}) P(k)+v_v e^{-(4+2 \epsilon)k r \pi}\left(v_v Q(k)-v_h (P(k)+Q(k)\right)\right] \nonumber \\
& +&\epsilon \left[v_h^2 \left(1+e^{-2(2+\epsilon)k r \pi}-e^{(2-2(2+\epsilon)) k r \pi}-2 e^{(2-4(2+\epsilon)) k r \pi}\right)-2 v_v v_h e^{-(4+\epsilon)k r \pi}\left(1-e^{(2-2(2+\epsilon)) k r \pi}\right)\right] \nonumber \\
&+&\left(\frac{2+\epsilon}{3+4 \epsilon}\right) v_v^2 e^{-2(4+\epsilon)k r \pi}\left[1+\omega^2 \left(F(k)+Q(k)\right)\right]+v_h^2e^{-(4+2 \epsilon)k r \pi}\left[1+\omega^2 \left(P(k)-G(k)\right)\right] \nonumber \\
&-&
v_v v_h e^{-(4+2 \epsilon)k r \pi}\left[1+\omega^2 (P(k)+F(k))+e^{-(4+2 \epsilon) k r \pi}(1+\omega^2 (Q(k)-G(k))\right]
\eea
where terms of order $\epsilon^2$ are neglected but $\epsilon k r$ is not treated as small. The quantities P, Q, F and G are given by  
\bea 
2 P(k)& = & \left(\frac{2 \nu}{1-\nu^2}\right) +\left(\frac{2+\nu}{1+\nu}\right)-\left(\frac{2(2-\nu^2)}{1-\nu^2}\right) e^{-(2+2 \epsilon)k r \pi} \\
2 Q(k) &= & \left(\frac{2 \nu}{1-\nu}\right)+\left(\frac{2+\nu}{1+\nu}\right)e^{2 k r \pi} \\
2 F(k) &=& \left(\frac{2+\nu}{1+\nu}\right)e^{2 k r \pi} \\
2 G(k) & = & \left(\frac{2 \nu}{1-\nu^2}\right) -\left(\frac{2+\nu}{1+\nu}\right) +\left(\frac{2(2-\nu^2)}{1-\nu^2}\right)e^{2 k r \pi}
\eea

Ignoring the terms proportional to $\epsilon$ we obtain the following conditions for which one achieve the minima of the potential
\bea
\label{eqn1}
4(v_v-v_h e^{-\epsilon k r \pi})-\frac{\omega^2}{3}e^{2 k r \pi}(4 v_v-11 v_h  e^{-\epsilon k r \pi}) = 0
\eea

It may be clearly seen from the equation (\ref{eqn1}) that for $\omega = 0$, the potential has a minima at 
\beq
\label{soln1}
k r=\frac{1}{\epsilon \pi} \ln (v_h/v_v)
\eeq 
which is identical to the solution of the Goldberger and Wise. Note that due to the 
presence of a non-zero brane cosmological constant the modified minima is obtained from  (\ref{eqn1}) as
\beq
\label{soln2}
k r=\frac{1}{\epsilon \pi} \ln (v_h/v_v) -\frac{7 \omega^2}{12 \epsilon \pi} (v_h/v_v)^{2/\epsilon}
\eeq

This result brings out an interesting feature. 
We know for $\omega = 0$ ( {\em i.e.} RS case ), a suitable non-hierarchical choice for 
$(v_h/v_v $ yields a stable value of $kr$ which is same as that proposed by Randall and sundrum to
resolve the gauge hierarchy problem. However  
From the figure (1), it can be seen that as we move into the Anti-de-Sitter
region from $\omega^2 =0$ ( i.e. RS case ) by increasing the magnitude of $\omega^2$ slightly , 
the required value of kr ( to resolve the gauge hierarchy problem) increases from RS value. On the other hand 
the corresponding stabilized value of $ k r$ decreases with increase in $\omega^2$, as has been shown in  (\ref{soln2}).
This opposite behavior clearly indicates that to achieve modulus stabilization and resolution of gauge hierarchy problem
simultaneously the most favored value of the brane cosmological constant $\omega = 0$ which was proposed in the original
model.

The entire stability analysis done so far is based on the AdS brane background. If one 
performs the same study with dS brane 
warp factor given in (\ref{wfds}) we end up with the conditions, 
\beq
\label{eqn3}
4(v_v-v_h e^{-\epsilon k r \pi})+ \frac{\omega^2}{3}e^{2 k r \pi}(4 v_v-11 v_h  e^{-\epsilon k r \pi}) = 0.
\eeq
with the stable modulus value as,
\beq
\label{soln3}
k r=\frac{1}{\epsilon \pi} \ln (v_h/v_v) +\frac{7 \omega^2}{12 \epsilon \pi} (v_h/v_v)^{2/\epsilon}
\eeq
Once again the figure(\ref{ADSDS}) and (\ref{soln3})reveals that while 
the required value of $kr$ which resolves the gauge hierarchy problem, decreases
with increase in the value of positive brane cosmological constant $\omega^2$, the corresponding stabilized value increases 
with $\omega^2$. Thus once again RS assumption of $\omega^2 = 0$ turns out to be a most favored solution. 

\section{conclusion} 

We now summarize the results obtained in this work. We have shown that for a generalized RS model with a cosmological 
constant on the brane, the modulus stabilization 
condition explicitly depends on the brane cosmological constant. We have obtained this 
condition both for positive and negative values of the brane cosmological constant. 
It turns out that if one increases the magnitude  of $\omega^2$ from zero the stabilized value of $k r$ decreases (increases)
for AdS (dS) brane. Whereas 
the value of $k r$ which solves the hierarchy problem increases (decreases) with the increase in value of the cosmological constant.
It implies that if one starts from $\omega^2 \sim 0$, then the variation of $k r$ with $\omega^2$  
for the requirement modulus stabilization and that for the gauge 
hierarchy resolution are opposite. This establishes that while the generalized RS model indeed predicts that for every value of brane
cosmological constant we have a corresponding value of $kr$ which can resolve the problem of gauge hierarchy, the
zero value of the brane cosmological constant, as assumed in the original RS model is most favored if
one imposes in addition the condition of modulus stabilization. The present small value of $\omega^2 \sim 10^{-124}$
yields the stabilized value of $kr$ for the de-Sitter case ( see equ. 3.20), which is very close to the corresponding value of $kr$
needed to resolve the gauge hierarchy problem ( see Figure 1 ). However for 
larger value of  $\omega^2$ the stabilized value of $kr$ is further away from that required to resolve the gauge hierarchy problem.
Thus a non-zero cosmological constant on the brane can consistently address both the gauge hierarchy as well as the modulus stabilization
problem only when it's value is extremely tiny. This is in agreement with the anthropic explanation \cite{anthro} in favour of
a small cosmological constant of our universe.

\acknowledgements{ JM acknowledges  Council For Scientific and Industrial Research, Govt. Of India for providing financial support. RK
acknowledges Department of Science and Technology, Govt. of India for providing financial support under the project grant 
No. SR/FTP/PS-68/2007}.

\end{document}